\documentclass[12pt]{article}

 \def\lddots{\mathinner{\mkern1mu\raise1pt\hbox{.}\mkern2mu

\raise4pt\hbox{.}\mkern2mu\raise7pt\vbox{\kern7pt\hbox{.}}\mkern1mu}}

\makeatletter

\def\numberbysection{\@addtoreset{equation}{section}
 \def\theequation{\thesection.\arabic{equation}}}

\makeatother

\numberbysection

\newcommand{\be}{\begin{eqnarray}}

\newcommand{\ee}{\end{eqnarray}}

\newcommand{\non}{\nonumber}

\begin{document}


\strut\hfill


\begin{center}

\LARGE  Thermodynamics of the critical $RSOS(q_{1}, q_{2} ;q)$ model\\[0.8in]

\large  Anastasia Doikou \footnote{e-mail: ad22@york.ac.uk} \\

\normalsize{ Department of Mathematics, University of York,
Heslington\\ York YO10 5DD, United Kingdom}\\

\end{center}


\begin{abstract}

The thermodynamic Bethe ansatz method is employed for the study of
the integrable critical $RSOS(q_{1}, q_{2};q)$ model. The high and
low temperature behavior are investigated, and the central charge
of the effective conformal field theory is derived. The obtained
central charge is expressed as the sum of the central charges of
two generalized coset models.

\end{abstract}


\section{Introduction}

It is well known that statistical systems at criticality ---second
order phase transition--- are expected to exhibit conformal
invariance \cite{pol}, therefore the critical behavior of such
systems should be described by a certain conformal field theory.
Different types of critical behavior have been classified
\cite{bpz}, and the critical exponents and correlation functions
have been determined (see also \cite{d}, \cite{fqs}).

An intriguing situation arises from the study of integrable
lattice models, whose scaling limit may correspond to certain
conformal field theories. In this framework an important, but non
trivial task is the calculation of the central charge of the
corresponding conformal field theory. A way one can extract this
information is by studying the finite size effects of the ground
state of the system \cite{VK}--\cite{fy}. An alternative approach
to compute the conformal properties is by investigating the low
temperature thermodynamics; in particular, the low temperature
behavior of the free energy of a critical system is described by
\cite{bc}, \cite{af} \be { F(T)\over L} = {F_{0}\over L} - {\pi c
\over 6 u}T^{2}+\dots, ~~T\ll 1. \ee For integrable theories this
can be achieved by means of the thermodynamic Bethe ansatz
approach, which is a powerful technique that allows the
computation of such properties. The mathematical techniques used
for such computations go back to the original work of several people
\cite{yy}--\cite{ts}.
The method was further treated and extended to various lattice
\cite{ftw}--\cite{mn1} (for a review on TBA for lattice models see
e.g. \cite{takamon}) and continuum relativistic models
\cite{z2}--\cite{dd} yielding very important results.

The thermodynamic Bethe ansatz for relativistic models is somehow
the inverse of the Bethe ansatz technique for lattice models
\cite{bethe}--\cite{korepin}. In the usual Bethe ansatz approach
the starting point is the microscopic Hamiltonian, whose
diagonalization gives rise to the Bethe ansatz equations, the
spectrum, and the scattering information
---expressed via the $S$ matrix--- (see e.g. \cite{FT2}, \cite{FT1}). On
the other hand, in the integrable relativistic theories one
employs the scattering information as an input in order to derive
the thermodynamics of the theory \cite{z2}, \cite{z1}.

In this study the thermodynamics of the $RSOS(q_{1},q_{2};q)$ is
investigated and the effective conformal anomaly is derived.
In general, $RSOS$ models are worth studying because, as already
mentioned, their critical behavior may be described by some
effective conformal field theory, e.g. critical fused $RSOS$
models are related to generalized diagonal coset models
(``anti--ferromagnetic'' regime) or parafermionic theories
(``ferromagnetic'' regime) \cite{BR}.
 Furthermore, it has been shown \cite{rs} that critical $RSOS$ models, with proper inhomogeneities, provide lattice regularizations of massive or massless integrable quantum field theories \cite{rs}, which on the other hand can be thought as perturbations of conformal field theories \cite{zamo}.
 What makes the $RSOS(q_{1}, q_{2};,q)$ model in particular interesting is that it is a natural generalization of the $RSOS(p,q)$ model studied by Bazhanov and Reshetikhin \cite{BR} in as much as the alternating spin chain, introduced by de Vega and Woyanorovich \cite{VEWO}, is a generalization of the fused $XXZ$ spin chain \cite{LT}. Therefore, with this article the study of the thermodynamics of the fused
critical $RSOS$ models is completed.

 In \cite{BR} the $RSOS(p,q)$ model was studied, the effective central charge was found and,  in the `` anti--ferromagnetic'' regime, it turned out to be the one of the $SU(2)$ diagonal coset model ${\cal M}(p,\nu-2-p)$ (${\cal M}(q,p) \equiv {SU(2)_{q}\otimes
SU(2)_{p} \over SU(2)_{q+p}}$, where $SU(2)_{k}$ is the $SU(2)$
$WZW$ model at level $k$ \cite{gko}, \cite{dms}), whereas in the
``ferromagnetic'' regime it agreed with  the central charge of the
parafermionic ${SU(2)_{\nu-2} \over U(1)}$ theory. In this work
the  effective central charge of the $RSOS(q_{1},q_{2};q)$ model is
computed from the low temperature analysis. In the
``anti--ferromagnetic'' regime it is expressed as the sum of the
central charges of two generalized diagonal coset models, namely
${\cal M}(q_{2},\nu -q_{2} -2)$ and ${\cal M}(q_{2},\delta q)$,
while in the ``ferromagnetic'' regime the analysis is exactly the
same as in \cite{BR}.

The outline of this article is as follows: in the next section
the model is introduced, and the Bethe ansatz equations and the
energy spectrum are presented. In the third section the
thermodynamic Bethe ansatz equations are derived explicitly and
the high and low temperature behavior are examined. Finally, from
the low temperature expansion the effective central charge is
derived.

\section{The model}

The integrable critical $RSOS(q_{1}, q_{2}; q)$ model, obtained
from the $RSOS(1,1)$ model by fusion \cite{KR}, \cite{djmo}, is
introduced. To describe the model, a square lattice of $2N$
horizontal and $M$ vertical sites is considered. The Boltzmann
weights  associated with every site are defined as
\begin{equation}
w(l_{i},l_{j},l_{m},l_{n}\vert \lambda)\equiv \left(
           \begin{array}{cc}
             l_{n}   &l_{m}         \\
              l_{i}   & l_{j}      \\
                                                         \end{array}\right)\,.\non\\
\label{boltzmann}
\end{equation}
With every face
$i$ of the lattice an integer $l_{i}$ is associated, and every
pair of adjacent integers satisfy the following restriction
conditions \cite{baxter}, \cite{abf} \be 0\leq l_{i+1} - l_{i} +P \leq 2P, (a) \non\\ P \leq l_{i+1} + l_{i} \leq 2\nu -P, (b) \label{condition} \ee where
$P=q_{1}$ for $i$ odd and $P=q_{2}$ for $i$ even (let $q_{1} >
q_{2}$), for the horizontal pairs, and $P=q$ for the vertical
pairs (array type II \cite{rs}).

 The fused Boltzmann weights
have been derived by Date $\it etal$ in \cite{djmo} and they are
given by
\be
w^{q_{i},1}(a_{1}, a_{q_{i}+1},b_{q_{i}+1},b_{1}\vert \lambda)
=\sum_{a_{2} \dots a_{q_{i}}} \prod_{k=1}^{q_{i}}w^{1,1}(a_{k},
a_{k+1},b_{k+1},b_{k}\vert \lambda +i(k-q_{i}))
\ee
where $b_{2} \dots b_{q_{i}}$ are arbitrary numbers satisfying
$\vert b_{i} -b_{i+1} \vert =1$. $w^{1,1}$ are the Boltzmann weights for the $SOS(1,1)$ model \cite{baxter}, they are
non vanishing as long as the condition (\ref{condition}(a)), for $P=1$ is
satisfied and they are given by the following expressions \be
w(l,l\pm 1,l,l\mp 1 \vert \lambda)=h(i-\lambda)  \non\\
w(l\pm 1,l,l\mp 1,l \vert \lambda)=-h(\lambda){h_{l+1} \over h_{l}}\non\\
w(l\pm 1,l,l\pm 1,l \vert \lambda)=h(w_{l}\pm \lambda){h_{1}
\over h_{l}} \label{boltzmann2} \ee where, \begin{equation}
h(\lambda) =\rho \Theta(\lambda)H(\lambda)
\end{equation} $H(\lambda)$ and $\Theta(\lambda)$ are Jacobi theta functions and,
\begin{equation} h_{l} =h(w_{l}), ~~~w_{l} =w_{0}+il. \end{equation}
 We are interested in the critical case where $h(\lambda)$ becomes a simple trigonometric function i.e., \be
h(\lambda) = {\sinh \mu \lambda \over \sin \mu}, \ee
 $w_{0}$,
$\rho$ and $\mu$ are arbitrary constants.
 Furthermore,
\be w^{q_{i}, q}(a_{1},
b_{1},b_{q+1},a_{q+1})=\prod_{k=0}^{q-2}\prod_{j=0}^{q_{i}-1} \Big
( h( i(k-j)+\lambda) \Big )^{-1} \non\\ \sum_{a_{2} \ldots a_{q}}
\prod_{k=1}^{q}w^{q_{i},1}(a_{k}, b_{k},b_{k+1},a_{k+1}\vert
\lambda+i (k-1)), \ee again $b_{2} \dots b_{q_{i}}$ are arbitrary
numbers satisfying $\vert b_{i} -b_{i+1} \vert =1$, and the pairs
$a_{1}$, $a_{q+1}$ and $b_{1}$,  $b_{q+1}$ satisfy
(\ref{condition}), for $P=q$. The fused weights satisfy the
Yang--Baxter equation in the following form \be \sum_{g}
w^{pq}(a,b,g,f\vert \lambda)w^{ps}(f,g,d,e\vert
\lambda+\mu)w^{qs}(g,b,c,d\vert \mu) \non\\ = \sum_{g}
w^{qs}(f,a,g,e\vert \mu)w^{ps}(a,b,c,g\vert \lambda+
\mu)w^{pq}(g,c,d,e\vert \lambda). \label{yb2} \ee Here we only
need the explicit expressions for $w^{q_{i},1}$ which are
 \be
w^{q_{i},1}(l+1,l'+1,l',l \vert
\lambda))=h_{q_{i}-1}^{q_{i}-1}(-\lambda)h_{a}
{h(ib-\lambda)\over h_{l}}  \non\\
w^{q_{i},1}(l+ 1,l'-1,l',l \vert \lambda))=h_{q_{i}-1}^{q_{i}-1}(-\lambda)h_{b}{h(\lambda+ia)
\over h_{l}}\non\\
w^{q_{i},1}(l-1,l'+1,l',l \vert
\lambda))=h_{q_{i}-1}^{q_{i}-1}(-\lambda)h_{c}{h(id-\lambda) \over h_{l}}\non\\
w^{q_{i},1}(l- 1,l'-1,l',l \vert
\lambda))=h_{q_{i}-1}^{q_{i}-1}(-\lambda)h_{d}{h(ic-\lambda) \over h_{l}}
\label{boltzmann3} \ee where \be a={l+l'-q_{i} \over
2},~~~b={l'-l+q_{i} \over 2}, ~~~c={l-l'+q_{i} \over 2},
~~~d={l+l'+q_{i} \over 2}, \ee
and
\be
h_{k}^{q}(\lambda) = \prod_{j=0}^{q-1} h\Big(\lambda+i(k-j)\Big). \label{h}\ee
 It is obvious that $w^{q_{i},1}(a,b,c,d\vert \lambda)$ are periodic functions, because they involve only simple trigonometric functions (\ref{boltzmann3}), (\ref{h}) ($h(\lambda+i\nu) =-h(\lambda)$, $\nu = {\pi \over \mu}$), i.e.
 \be
 w^{q_{i},1}(a,b,c,d\vert \lambda +i\nu) = (-)^{q_{i}}w^{q_{i},1}(a,b,c,d\vert \lambda)
 \label{period} \ee

Now we can define the transfer matrix of the $RSOS(q_{1},q_{2};q)$
model \be T^{q_{1},q_{2};q\{b_{1}\ldots b_{2N} \}}_{\{a_{1} \ldots
a_{2N}\}} = \prod_{j=1}^{2N-1}w^{q_{1},q}(a_{j},a_{j+1},
b_{j+1},b_{j}\vert \lambda)w^{q_{2},q}(a_{j+1},a_{j+2},
b_{j+2},b_{j+1}\vert \lambda)\ee where we impose periodic boundary
conditions, i.e. $a_{2N+1} =a_{1}$ and $b_{2N+1} =b_{1}$. Notice
that in the odd and even sites the weights $w^{q_{1},q}$ and
$w^{q_{2},q}$ live respectively. The case where $q_{1} =q_{2}$
(array type I \cite{rs}), namely the fused $RSOS(p,q)$ model, has
been studied in detail by Bazhanov and Reshetikhin in \cite{BR}.
It is evident that the model studied here is a generalization of
the fused $RSOS(p,q)$ model. The analogue of the array type II in
the spin chain framework is the alternating quantum spin chain,
introduced by de Vega and Woyanorovich \cite{VEWO}, and also
studied extensively by many authors \cite{AM}--\cite{bydo}.

From the Yang--Baxter equation for the fused Boltzmann
weights (\ref{yb2}) the commutativity
property for the transfer matrix follows, i.e. \be
T^{q_{1},q_{2};q}(\lambda)T^{q_{1},q_{2};q'}(\mu)
=T^{q_{1},q_{2};q'}(\mu)T^{q_{1},q_{2};q}(\lambda),\label{commu1}
\ee
Moreover the transfer matrix is periodic (\ref{period})
\be
T^{q_{1},q_{2};q}(\lambda +i\nu) = T^{q_{1},q_{2};q}(\lambda). \label{period2}
\ee
In order to obtain the Bethe ansatz equations for the model we
also need the following useful relations. First we will use the
relations acquired by the fusion procedure \cite{djmo}, \cite{BR},
namely \be T_{0}^{q_{1},q_{2};q}T_{q}^{q_{1},q_{2};1} =
f_{q}^{q_{1},q_{2}}T_{0}^{q_{1},q_{2};q-1}+f_{q-1}^{q_{1},q_{2}}T_{0}^{q_{1},q_{2};q+1}
\label{fusion} \ee where \be f_{q}^{q_{1},q_{2}}(\lambda) = \Big
(h_{q}^{q_{1}}(\lambda) h_{q}^{q_{2}}(\lambda)\Big )^{N},
~~~T_{k}^{q_{1},q_{2};q} =T^{q_{1},q_{2};q}(\lambda+ik)
\label{rel2}, ~~T_{0}^{q_{1},q_{2};0} = f_{-1}^{q_{1},q_{2}}.
\label{fusion2}\ee Notice that the main difference between
equations (\ref{fusion}), (\ref{fusion2}) and the corresponding
equations in \cite{BR} is the substitution of $p$ with
$q_{1},q_{2}$. In particular $f_{q}^{p}$ in \cite{BR} is replaced
here by $f_{q}^{q_{1},q_{2}}$. We must also have in mind that the Boltzmann weights satisfy the
following important property, i.e. up to a gauge transformation,
that does not affect the transfer matrix, the weights
$w^{1,q}(a,b,c,d\vert \lambda)$ and
$w^{1,\nu-2-q}(\nu-a,\nu-b,c,d\vert \lambda+i(q+1))$ coincide, where \be
w^{1,q}(a,d,c,b\vert\lambda-i(q-1))=\Big
(h_{q-1}^{q-1}(-\lambda)\Big ) ^{-1}w^{q,1}(a,b,c,d\vert \lambda),
\ee a similar property holds also between the  weights
$w^{q_{i},q}$ and $w^{q_{i},\nu-2-q}$. From the above relations it
follows that
 \be
T^{q_{1},q_{2};q}(\lambda) &=& YT^{q_{1},q_{2};\nu-2-q}(\lambda+i(q+1)), ~~~q=1,\ldots,\nu-3, \non\\  T^{q_{1},q_{2};\nu -2}(\lambda) &=& Y\Big (h_{\nu-2}^{q_{1}}(\lambda)h_{\nu-2}^{q_{2}}(\lambda) \Big)^{N} \label{rel4}
\ee
with
\be Y_{\{ l_{1} \ldots
l_{2N} \} }^{\{ l'_{1} \ldots l'_{2N}\} }=\prod_{i=1}^{2N} \delta(l_{i},
\nu-l'_{i}), ~~\Big [T^{q_{1},q_{2};q},Y\Big] =0. \label{rel3} \ee

To derive the transfer matrix eigenvalues we employ the
commutativity properties of the transfer matrix (\ref{commu1}),
(\ref{rel3}), the periodicity (\ref{period}), (\ref{period2}), the
fusion relations (\ref{fusion}), (\ref{fusion2}), equations
(\ref{rel4}) and the analyticity of the eigenvalues. 
Moreover, we employ relations
(\ref{fusion}) and (\ref{rel4}) for $q=\nu-1, \nu$ and we derive
\be
T^{q_{1},q_{2};\nu -1}(\lambda) =0, ~~~ T^{q_{1},q_{2};\nu}(\lambda)=-Yf_{\nu-1}^{q_{1},q_{2}}(\lambda). \label{formal}
\ee
 From the
solution of the above system of equations (\ref{commu1})--(\ref{rel3}), and with the help of relations (\ref{formal}) we can write
equation (\ref{fusion}) in
the following form \be detM[\Lambda^{q_{1},q_{2};1}(\lambda)]=0
\ee where
\begin{equation}
M[\Lambda^{q_{1},q_{2};1}(\lambda)]= \left(
        \begin{array}{cccccccc}
 \Lambda_{0}^{q_{1},q_{2};1} &f_{-1}^{q_{1},q_{2}} &0 &0 &.&0 &0  &-Yf_{0}^{q_{1},q_{2}}    \\
 f_{1}^{q_{1},q_{2}} &\Lambda_{1}^{q_{1},q_{2};1} &f_{0}^{q_{1},q_{2}} &0 &.&0 &0 &0  \\
 0 &f_{2}^{q_{1},q_{2}} &\Lambda_{2}^{q_{1},q_{2};1} &f_{1}^{q_{1},q_{2}} &. &0 &0 &0  \\
 . &.   &.   &.    &.  &.   &.   &.  \\
 0 &0 &0 &0 &. &f_{\nu-2}^{q_{1},q_{2}} &\Lambda_{\nu-2}^{q_{1},q_{2};1} &f_{\nu-3}^{q_{1},q_{2}}  \\
-Yf_{\nu-2}^{q_{1},q_{2}} &0 &0 &0 &.&0 &f_{\nu-1}^{q_{1},q_{2}}&\Lambda_{\nu-1}^{q_{1},q_{2};1} \\                       \end{array}\right)\,.\non\\
\label{matrix2}
\end{equation}
Let now $(Q_{0}^{q_{1},q_{2}}(\lambda), \ldots ,Q_{\nu-1}^{q_{1},q_{2}}(\lambda))$ be the null vector of the matrix (\ref{matrix2}) with $Q_{k}^{q_{1},q_{2}}(\lambda)=\omega^{k}Q^{q_{1},q_{2}}(\lambda+ik)$, $\omega^{2\nu}=1$
and
\be
Q^{q_{1},q_{2}}(\lambda) = \prod_{j=1}^{{(q_{1}+q_{2})N
\over 2}}h(\lambda-\lambda_{j}),
 \ee
 then the eigenavlues  are given by the following expression
 \be \Lambda^{q_{1},q_{2};1}(\lambda) =
\omega f_{-1}^{q_{1},q_{2}}(\lambda) {Q^{q_{1},q_{2}}(\lambda
+i) \over Q^{q_{1},q_{2}}(\lambda)} +\omega^{-1}
f_{0}^{q_{1},q_{2}}(\lambda){Q^{q_{1},q_{2}}(\lambda-i) \over Q^{q_{1},q_{2}}(\lambda)}. \label{eigen} \ee
For completeness we write the general expression of the eigenvalues $\Lambda^{q_{1},q_{2};q}(\lambda)$, which follow from the fusion relation (\ref{fusion}) and (\ref{eigen}),
\be
\Lambda^{q_{1},q_{2};q}(\lambda) =
 Q^{q_{1},q_{2}}(\lambda -i) Q^{q_{1},q_{2}}(\lambda
+iq) \sum_{j=0}^{q} {\omega^{q-2j}f^{q_{1},q_{2}}(\lambda+i(j-1)) \over Q^{q_{1},q_{2}}(\lambda+i(j-1))Q^{q_{1},q_{2}}(\lambda+ij)}. \label{gener} \ee
The eigenvalues satisfy all equations (\ref{fusion}), (\ref{fusion2}) and (\ref{rel4}),
 where
$\omega$ is a root of unity that obeys the constraint \be
\omega^{\nu} = -(-)^{{(q_{1}+q_{2})N \over 2}}y \label{cons3} \ee
and $y=\pm 1$ is the eigenvalue of the operator $Y$ (\ref{rel3}).
Equation (\ref{cons3}) is a consequence of the periodicity and (\ref{rel4}).
 Similarly, here the difference with the corresponding eigenvalues in \cite{BR} is the replacement of the functions $f^{p}$ and $Q^{p}$ with $f^{q_{1},q_{2}}$ and $Q^{q_{1},q_{2}}$ respectively.
Finally, from the analyticity of the eigenvalues we obtain the Bethe ansatz equations 
\begin{equation}
\omega^{-2}e_{q_{1}}(\lambda_{\alpha})^{N}e_{q_{2}}(\lambda_{\alpha})^{N}=
-\prod_{\beta=1}^{M} e_2(\lambda_{\alpha}-\lambda_{\beta})
\label{BAE}
\end{equation}
where \begin{equation} e_{n}(\lambda; \nu)=\frac{\sinh
\mu(\lambda+{in\over 2})}{\sinh \mu(\lambda-{in \over 2})}.
\nonumber
\end{equation}
  It is important to emphasize that the eigenstates
of the model are states with zero spin $S_{z}=0$ \cite{BR},
\cite{ps}, \cite{rs}, i.e.
\begin{equation}
M={1\over 4}(q_{1} + q_{2})L \\, \label{spin}
\end{equation}
where $L=2N$ (for $q_{1} =q_{2} =p$ the later constraint agrees
with the corresponding constraint in \cite{BR}). We should mention
that the Bethe ansatz equations (\ref{BAE}) have the same
structure with the Bethe ansatz equations of the alternating
${q_{1} \over 2}, {q_{2} \over 2}$ spin chain
\cite{VEWO}--\cite{bado}. The main differences between the model
under study and the alternating spin chain are: 1) the phase
$\omega$ which is unit, and 2) the number of strings $M$ which is
not fixed in the alternating spin chain.

 The energy\footnote{The
Hamiltonian of the model is defined for $q=q_{1}, q_{2}$ \be
H=-{\mu\over 8\pi } \sum_{i=1}^{2}{d\over d\lambda}\ln
T^{q_{1},q_{2};q_{i}}(\lambda)\vert _{\lambda=0}, \ee
 where $T^{q_{1},q_{2};q_{i}}$ is the transfer matrix of the $RSOS(q_{1},q_{2};q_{i})$ model (see also (\ref{gener})).}
   of a
 state is
characterized by the set of quasi particles with rapidities (Bethe
ansatz roots) {$\lambda_{j}$}, \cite{FT2}, \cite{FT1}, \cite{KR},
\begin{equation}
E=-{\mu \over 8 \pi }\sum_{j=1}^{M} \sum_{n =1}^{2} { \sin \mu
q_{n} \over \sinh \mu (\lambda_{j}+ {iq_{n}\over 2})\sinh \mu
(\lambda_{j}- {iq_{n} \over 2})}. \label{energy}
\end{equation}
 The
thermodynamic limit $N \rightarrow \infty$ of the  equation
(\ref{BAE}) can be studied with the help of the string hypothesis
\cite{g}, \cite{t1},
\cite{FT2}, \cite{FT1}, which states that solutions of (\ref{BAE})
in the thermodynamic limit are grouped into strings of length $n$
with the same real part and equidistant imaginary parts \be
\lambda_{\alpha}^{(n,j)} &=&\lambda_{\alpha}^n + {i\over
2}(n+1-2j),~~j=1,2,...,n, \non\\
\lambda_{\alpha}^{(0,s)}&=&\lambda_{\alpha}^{0}+i\frac{\pi}{2\mu},
 \label{STR} \ee where $\lambda_{\alpha}^n$ and $\lambda_{\alpha}^{0}$ are real,
 and $\lambda_{\alpha}^{(0,s)}$ is the negative parity string.
The allowed strings that describe the thermodynamics of the model
are the same as in \cite{BR} and they are $ 1\leq n\leq \nu -2$ ($q_{i}\leq
\nu-2$), the negative parity string is also excluded. Then, the
Bethe ansatz equations (\ref{BAE}) following \cite{g}, \cite{t1}  become,

\begin{equation}
\omega^{-2}\prod _{j= 1}^{2}X_{nq_{j}}(\lambda_{\alpha}^n)^{N}=
-\prod_{m= 1}^{\nu -2} \prod_{\beta=1}^{M_{m}}
E_{nm}(\lambda_{\alpha}^{n}-\lambda_{\beta}^{m}) \label{K}
\end{equation}
where $n=1, \dots,\nu-2$, and
\begin{eqnarray}
X_{nm}(\lambda)&=& e_{|n-m+1|}(\lambda) e_{|n-m+3|}(\lambda)\ldots
e_{(n+m-3)}(\lambda)e_{(n+m-1)}(\lambda)
\nonumber\\
E_{nm}(\lambda)&=& e_{|n-m|}(\lambda)e_{|n-m+2|}^{2}(\lambda)
\ldots e_{(n+m-2)}^{2}(\lambda)e_{(n+m)}(\lambda).
\end{eqnarray}
Finally, the energy (\ref{energy}) by virtue of the string
hypothesis (\ref{STR}) takes the form \be E=-{L \over 4}
\sum_{n=1}^{\nu -2} \int_{-\infty}^{\infty}d\lambda
(Z_{nq_{1}}^{(\nu)}(\lambda)
+Z_{nq_{2}}^{(\nu)}(\lambda))\rho_{n}(\lambda) \label{energy2} \ee
where, $\rho_{n}$ is the density\footnote{here we use the
Maclaurin expansion \be \sum_{j=1}^{M} f(\lambda_{j}) \sim L
\int_{-\infty}^{\infty} f(\lambda) \rho(\lambda)d\lambda,
\label{mac} \ee} of the $n$ strings (pseudo-particles) and
\begin{equation}
Z_{nm}^{(\nu)}(\lambda)= {1 \over 2\pi}{d \over d\lambda} i \log
X_{nm}(\lambda)\,,
\end{equation}
the Fourier transform of the last expression is
\begin{equation}
\hat Z_{nm}^{(\nu)}(\omega)= { \sinh \Bigl (( \nu -
\max(n,m)){\omega \over 2} \Bigr ) \sinh \Bigl (\min(n,m){\omega
\over 2} \Bigr ) \over \sinh ({\nu \omega \over 2}) \sinh({\omega
\over 2})}\\.
\end{equation}

\section{Thermodynamic Bethe Ansatz}

In what follows the thermodynamic Bethe ansatz equations are
derived from (\ref{K}). In addition to the density of
pseudo--particles $\rho_{n}$ we also introduce the density of
holes $\tilde \rho_{n}$, and we can immediately deduce from
(\ref{K}), and with the help of the Maclaurin expansion
(\ref{mac}) that they satisfy
 \be  \tilde \rho_{n}(\lambda)=
{1\over 2}(Z_{nq_{1}}^{(\nu)}(\lambda )+
Z_{nq_{2}}^{(\nu)}(\lambda)) - \sum_{m=1}^{\nu -2}A_{nm}^{(\nu)}
* \rho_{m}(\lambda).
\label{densities} \ee where
\begin{equation}
\ A_{nm}^{(\nu)}(\lambda)= {1 \over 2\pi}{d \over d\lambda} i \log
E_{nm}(\lambda)+\delta_{nm}\delta(\lambda)\,,
\end{equation} and
\begin{equation}
\hat A_{nm}^{(\nu)}(\omega)={2 \coth ({\omega \over 2} )\sinh
\Bigl (( \nu - \max(n,m)){\omega \over 2}\Bigr) \sinh \Bigl
(\min(n,m){\omega \over 2}\Bigr ) \over \sinh ({\nu \omega \over
2})}\\.
\end{equation}
 However, recall that the only allowed states as in \cite{BR} are the ones
with $S_{z} =0$ and therefore from
(\ref{spin}),   \be \sum_{n=1}^{\nu -2}n \int_{-\infty}^{\infty}
\rho_{n}(\lambda)d\lambda = {q_{1} +q_{2} \over 4}. \label{constr1} \ee  Equation (\ref{constr1}) together with relation (\ref{densities}) for $n=\nu-2$
yields \be \int_{-\infty}^{\infty} \tilde \rho_{\nu
-2}(\lambda)d\lambda =0 \Rightarrow \tilde \rho_{\nu
-2}(\lambda)=0. \label{c1} \ee The constraint (\ref{c1}) is
imposed on (\ref{densities}) and the density $\rho_{\nu-2}$ is
expressed in terms of the rest densities,
 \be
\rho_{\nu -2}(\lambda) = \rho^{0}(\lambda)-\sum_{m=1}^{\nu-3}
a_{\nu -2 -m}^{(\nu -2)}*\rho_{m}(\lambda) \label{ro} \ee where
\be \hat a_{n}^{(\nu-2)}(\omega)= {\sinh \Big ((\nu -n -2){\omega
\over 2}\Big) \over \sinh\Big ((\nu-2){\omega \over 2}\Big)}, ~~
\hat \rho^{0}(\omega) = {\sinh (q_{1}{ \omega \over 2}) +\sinh
(q_{2}{ \omega \over 2}) \over 4 \cosh({\omega \over 2})
\sinh((\nu -2){\omega \over 2})}.
 \ee By means of the relation (\ref{ro}) the equation
(\ref{densities}) can be rewritten in the following form \be
\tilde \rho_{n}(\lambda)= {1\over 2}(Z_{nq_{1}}^{(\nu
-2)}(\lambda)+ Z_{nq_{2}}^{(\nu -2)}(\lambda)) - \sum_{m=1}^{\nu
-3}A_{nm}^{(\nu -2)}
* \rho_{m}(\lambda). \label{densities2} \ee

 The energy of the system, after we
apply the string hypothesis is given by (\ref{energy2}). Now,
taking into account the equation (\ref{densities2}) the energy
becomes \be e={E \over L} = -g_{0} -{1\over 4}\sum_{n=1}^{\nu -3}
\int_{-\infty}^{\infty}d\lambda (Z_{nq_{1}}^{(\nu -2)}(\lambda)
+Z_{nq_{2}}^{(\nu -2)}(\lambda))\rho_{n}(\lambda) \label{energy3}
\ee with \be g_{0} = {1\over 16 \pi}\int_{-\infty}^{\infty}
d\omega {\Big (\sinh (q_{1}{\omega \over 2}) +\sinh (q_{2}{\omega
\over 2})\Big )^{2} \over \sinh ({\nu \omega \over 2})\sinh ((\nu
-2){\omega \over 2})}. \ee

In order to determine the thermodynamic Bethe ansatz equations the
free energy of the system should be minimized, i.e., $\delta F
=0$, where \be F=E-TS, \label{f} \ee and the entropy of the system
is given by, \be S &\simeq& L \sum_{n=1}^{\nu -3}
\int_{-\infty}^{\infty} d \lambda \Big ((\rho_{n}(\lambda)+\tilde
\rho_{n}(\lambda)) \ln(\rho_{n}(\lambda)+\tilde \rho_{n}(\lambda))
-\rho_{n}(\lambda) \ln \rho_{n}(\lambda)- \tilde \rho_{n}(\lambda)
\ln \tilde \rho_{n}(\lambda)\Big )\non\\ &=& L\sum_{n=1}^{\nu -3}
 \int_{-\infty}^{\infty} d \lambda \Big (\rho_{n}(\lambda)
\ln(1+{\tilde \rho_{n}(\lambda) \over \rho_{n}(\lambda)}) + \tilde
\rho_{n}(\lambda) \ln(1+{ \rho_{n}(\lambda) \over \tilde
\rho_{n}(\lambda)})\Big ). \label{entropy}\ee Then, from equations
(\ref{energy3}), (\ref{entropy}) and the constraint
(\ref{densities2}) the following expression is implied \be T\ln
\Big (1+\eta_{n}(\lambda)\Big ) = -{1\over 4}(Z_{nq_{1}}^{(\nu
-2)}(\lambda)+Z_{nq_{2}}^{(\nu -2)}(\lambda)) + \sum_{m=1}^{\nu
-3} A_{nm}^{(\nu -2)}* T\ln \Big (1+\eta_{m}^{-1}(\lambda)\Big ),
\label{TBA} \ee where $\eta_{n}(\lambda) ={\tilde
\rho_{n}(\lambda) \over \rho_{n}(\lambda)}$. It is convenient to
consider the convolution of the expression (\ref{TBA}) with the
inverse of $A_{nm}$, \be \hat A_{nm}^{-1}(\omega) = \delta_{nm}
-\hat s(\omega)(\delta_{nm+1} +\delta_{nm-1}), \ee having in mind
the following identity, \be A_{nm}^{-1} * Z_{mq_{i}}(\lambda) =
s(\lambda) \delta_{nq_{i}}, \ee where \be s(\lambda) = {1\over 2
\cosh (\pi \lambda)},~~\hat s(\omega) ={1\over 2\cosh({\omega
\over 2})}, \ee and $\eta_{n}(\lambda) =e^{{\epsilon_{n}(\lambda)
\over T}}$, (\ref{TBA}) becomes, \be \epsilon_{n}(\lambda)
&=&s(\lambda)*T \ln (1+ \eta_{n+1}(\lambda))(1+
\eta_{n-1}(\lambda)) -{1\over 4}s(\lambda)(\delta_{nq_{1}}
+\delta_{nq_{2}}), \label{TBA2} \ee for any $n=1, \ldots, \nu-3$.
Note that the last equation differs from the corresponding equation obtained in \cite{BR} in
 the inhomogeneity term $s(\lambda)$. More specifically, here the terms $\delta_{nq_{1}}$
and $\delta_{nq_{2}}$ appear, whereas in the study of the fused
$RSOS(p,q)$ model \cite{BR} only the $\delta_{np}$ term appears.
It is obvious that for $q_{1}=q_{2}=p$ our expression agrees with
the corresponding expression for the pseudo--energies in
\cite{BR}. It can be easily deduced from equation (\ref{TBA2})
that the pseudo--energy $\epsilon_{n}(\lambda)>0$ for every $n
\neq q_{1}, q_{2}$, therefore we conclude that the ground state
consists of two filled Dirac seas with strings of length $q_{1},
q_{2}$, i.e. $\tilde \rho_{n}(\lambda) =0$ for any $n$, and $
\rho_{n}(\lambda) =0$ for any $n\neq q_{1}, q_{2}$. The
pseudo--energies for those are immediately induced from
(\ref{TBA}) by neglecting the terms of the sum for $m\neq q_{i}$,
\be \epsilon_{i}(\lambda) = -{1\over 4}
\sum_{j=1}^{2}Z_{q_{i}q_{j}}^{(\nu-2)}(\lambda) +
\sum_{j=1}^{2}\tilde A_{q_{i}q_{j}}^{(\nu-2)}* T
\ln(1+\eta_{q_{j}}^{-1}(\lambda)),~~i=1,2
\label{appr} \ee (N.B.
$\epsilon_{i}(\lambda)\equiv\epsilon_{q_{i}}(\lambda)$) where \be
\tilde A_{nm}^{(\nu-2)}(\lambda) = A_{nm}^{(\nu-2)}(\lambda)
-\delta_{nm} \delta(\lambda). \ee
Moreover, the energy of the ground state can be written from
(\ref{densities2}), (\ref{energy3}) \be e_{0} &=&{E_{0}\over L}=
-g_{0} -{1\over 8}\sum_{i, j=1}^{2}\int_{-\infty}^{\infty}d\lambda
Z_{q_{i}q_{j}}^{(\nu -2 )}(\lambda)s(\lambda) \non\\
&=& - {1\over 8}\sum_{i, j=1}^{2}\int_{-\infty}^{\infty}d\lambda
Z_{q_{i}q_{j}}^{(\nu)}(\lambda)s(\lambda). \label{energy4} \ee The
free energy of the system follows from (\ref{energy3}), (\ref{f}),
(\ref{entropy}), (\ref{densities2}), and (\ref{TBA}), \be
f(T)={F(T) \over L}= -g_{0}- {T\over 2}\sum_{n=1}^{\nu-3}
\int_{-\infty}^{\infty} d \lambda
\ln(1+\eta_{n}^{-1}(\lambda))(Z_{nq_{1}}^{(\nu -2)}(\lambda)
+Z_{nq_{2}}^{(\nu -2)}(\lambda)), \label{fe}\ee and in terms of
the ground state energy of the system (\ref{energy4}) we can write
\be f(T)= e_{0}- {T \over 2}\sum_{i=1}^{2}\int_{-\infty}^{\infty}
d \lambda s(\lambda)\ln(1+\eta_{q_{i}}(\lambda)). \label{free1}
\ee In the following sections we are going to explore the behavior
of the free energy and the entropy of the system in the high and
low temperature.

\subsection{The high temperature expansion}

By studying the high temperature behavior of the entropy the
number of states of the model can be deduced. In the high
temperature limit the pseudo--energies $\epsilon_{n}$ become
independent of $\lambda$ \cite{BT}, consequently the thermodynamic
Bethe ansatz equations (\ref{TBA2}) are given by \be \epsilon_{n}
&\simeq& s(\lambda)*T \ln (1+ \eta_{n+1})(1+ \eta_{n-1}) \non\\
&=& {T \over 2} \ln (1+ \eta_{n+1})(1+ \eta_{n-1}), \label{TBA3}
\ee  and the corresponding solution of the above difference
equation is exactly the same as in  \cite{BR} (for $T \to \infty$ the inhomogeneity
 term can be neglected in (\ref{TBA2}) and therefore the pseudo--energies coincide with the ones found in \cite{BR}) \be \ln (1 + \eta_{n}) = \ln
{\sin^{2} ({\pi (n+1) \over \nu}) \over \sin^{2} ({\pi \over
\nu})}. \label{TBA4} \ee The free energy follows immediately from
(\ref{free1}), (\ref{TBA4}) \be F=-{TL \over 4} \sum_{n=q_{1},
q_{2}}\ln {\sin^{2} ({\pi (n+1) \over \nu}) \over \sin^{2} ({\pi
\over \nu})}, \ee moreover, the entropy in the high temperature
limit (\ref{f}) becomes \be S= {L \over 2}\sum_{n=q_{1}, q_{2}}
\ln {\sin ({\pi (n+1) \over \nu}) \over \sin ({\pi \over \nu})}.
\label{s}\ee Notice here that the free energy and the entropy are
expressed as a sum of two terms since the ground state consists of
two filled Dirac seas. On the other hand, in \cite{BR} the
corresponding expressions contain just one term, because the
ground state there consists of one filled Dirac sea. Finally, we
conclude that the number of states for the system is
 \be
\prod_{n=q_{1}, q_{2}} \Big ( {\sin ({\pi (n+1) \over \nu}) \over
\sin ({\pi \over \nu})}\Big )^{{L\over 2}}. \ee Notice that in the
isotropic limit $\nu \to \infty$ the entropy (\ref{s}) coincides
with the one of the alternating ${q_{1} \over 2}$, ${q_{2} \over
2}$ spin chain (see e.g. \cite{DMN1}, \cite{bydo}). For $q_{1} =q_{2}$ (\ref{s}) agrees with the entropy found in \cite{BR}.

\subsection{The low temperature expansion}

The main purpose of this section is the derivation of the
effective central charge via the study of the low temperature
thermodynamics. Recall, that the ground state of the model
consists of two filled Dirac seas of strings $q_{1}, q_{2}$,
therefore we examine the TBA (\ref{TBA}) for $n=q_{1}, q_{2}$. In
the $T\to 0$ limit the following quantities are defined \be T
\ln(1+\eta_{i}^{\pm}) \to \pm \epsilon_{i}^{\pm}, ~~i=1,2 \ee
with, \be \epsilon_{i}^{-} = {1\over 2}(\epsilon_{i}
-|\epsilon_{i}|), ~~\epsilon_{i}^{+} = \epsilon_{i}
-\epsilon_{i}^{-}, \ee then the pseudo-energies for the ground
state (\ref{appr}) take the form \be \epsilon_{i}(\lambda) =
-{1\over 4} \sum_{j=1}^{2}Z_{q_{i}q_{j}}^{(\nu-2)}(\lambda)
 -\sum_{j=1}^{2}\tilde
 A_{q_{i}q_{j}}^{(\nu-2)}*\epsilon_{j}^{-}(\lambda).
\ee Finally, the last equation can be written in terms of
$\epsilon_{i}, \epsilon_{i}^{+}$ \be
\sum_{j=1}^{2}A_{q_{i}q_{j}}^{(\nu-2)} *\epsilon_{j}(\lambda) =
-{1\over 4}\sum_{j=1}^{2}Z_{q_{i}q_{j}}^{(\nu-2)}(\lambda) +
\sum_{j=1}^{2}\tilde A_{q_{i}q_{j}}^{(\nu-2)}*
\epsilon_{j}^{+}(\lambda) \label{appr2}, \ee and the solution of
the above system is given by the following expression \be
\epsilon_{i}(\lambda) = -{1\over 4}s(\lambda) +
\sum_{j=1}^{2}K_{ij}* \epsilon_{j}^{+}(\lambda),~~i=1,2
\ee where the kernel $K$ is \be K(\lambda)=\left(
           \begin{array}{cc}
             h_{1}(\lambda) & h(\lambda)         \\
             h(\lambda)     & h_{2}(\lambda)\\
                                                         \end{array}\right)\,
                                                        \label{matrix},   \ee
\be \hat h_{1}(\omega) &=& {\sinh ((\delta q -1){\omega \over 2})
\over 2\cosh({\omega \over 2})\sinh(\delta q{\omega \over 2})}
+{\sinh ((\nu -3 -q_{1}){\omega \over 2}) \over 2 \cosh({\omega
\over 2})\sinh((\nu -2 -q_{1}){\omega \over 2})}, \non\\ \hat
h_{2}(\omega) &=&{\sinh ((\delta q -1){\omega \over 2}) \over
2\cosh({\omega \over 2})\sinh (\delta q{\omega \over 2})}+ {\sinh
((q_{2}-1){\omega \over 2}) \over 2 \cosh({\omega \over 2})\sinh
(q_{2}{\omega \over 2})},~~\hat h(\omega) = {\sinh ({\omega \over
2}) \over 2\cosh({\omega \over 2})\sinh (\delta q{\omega \over
2})} \label{hh}\ee and $\delta q = q_{1} -q_{2}$. Note, that the expression of the kernel (\ref{matrix}), (\ref{hh}) in this general form for any $q_{1}$, $q_{2}$ is  rather a new result. As
long as the condition $q_{1} =\nu-2-q_{2}$ holds,
the symmetry between left and right sectors is satisfied (see also
e.g. \cite{rs}). In particular, $h_{1} = h_{2}$, with $h_{1},
h_{2}$ being related to the scattering in the left (right) sector.
In general, for $\delta q\neq 1$ each of $h_{i}$ is decomposed
into two parts (see (\ref{hh})), and every part is related to the
triplet amplitude of the $XXZ$ model, with different anisotropy
parameters (hidden degrees of freedom \cite{pw}, \cite{KR},
\cite{bydo}). In the special case where $\delta q=1$, there are no
hidden degrees of freedom, and $h_{1}, h_{2}$ are relevant to the
triplet amplitudes of the $XXZ$ (sine--Gordon) model with the
proper anisotropy parameters, whereas $h$ corresponds to the
massless LR scattering amplitude (see also \cite{zz},
\cite{bado}).

To derive the effective central charge, the
entropy of the system must be evaluated in the low temperature
limit. In order to do that the following approximations, which
hold true for $\lambda \to \infty$, should be made \cite{ftw},
\cite{B}, \cite{BT}, \be \rho_{n}(\lambda) \simeq{2 \over
\pi}f_{n}(\lambda) {d \over d \lambda} \epsilon_{n}(\lambda),
~~\tilde \rho_{n}(\lambda) \simeq {2 \over \pi}
(1-f_{n}(\lambda)){d \over d \lambda} \epsilon_{n}(\lambda) \ee
where $f_{n}(\lambda) = (1+e^{{\epsilon_{n}(\lambda) \over
T}})^{-1}$, ($f_{0}(\lambda) =f_{\nu-2}(\lambda)\equiv1$), and the
entropy (\ref{entropy}), can be written as \be s={S \over L} = -{2
\over \pi} \sum_{n=1}^{\nu-3}
\int_{\epsilon_{n}(-\infty)}^{\epsilon_{n}(\infty)}d\epsilon_{n}
\Big (f_{n}(\lambda) \ln f_{n}(\lambda) +(1-f_{n}(\lambda)) \ln
(1-f_{n}(\lambda))\Big ). \ee By changing variables in the last
expression, \be s = {2 T\over \pi}  \sum_{n=1}^{\nu-3}
\int_{f_{n}^{min}}^{f_{n}^{max}}df_{n} \Big ( {\ln f_{n} \over
1-f_{n}} +{ \ln (1-f_{n}) \over f_{n}} \Big ) , \ee and by
introducing the Rogers dilogarithm  \be L(x) =-{1\over 2}
\int_{0}^{x} dy\Big ({\ln y \over 1-y} + {\ln (1-y) \over y}\Big )
\ee the entropy can be written in terms of the dilogarithms  as
follows \be s = -{4T \over \pi} \sum_{n=1}^{\nu -3} \Big
(L(f_{n}^{max})-L(f_{n}^{min})\Big). \label{entropy1}\ee

The next natural step is the solution of the TBA equations
(\ref{TBA2}) in the low temperature limit. In order to do that it
is convenient (see also \cite{ftw}, \cite{B}, \cite{BT},
\cite{BR}) to introduce the function \be \phi_{n}(\lambda) =
{1\over T} \epsilon_{n}(\lambda-{1\over \pi} \ln T ), \ee then the
TBA equations become , \be \phi_{n} \simeq -s(\lambda)* \ln
f_{n+1}f_{n-1}-{1\over 4}e^{-\pi \lambda}(\delta_{nq_{1}}
+\delta_{nq_{2}}) \label{differ}. \ee Our task is to solve the
later difference equation in the limit that $\lambda \to \pm
\infty$, ($\phi_{n}$ independent of $\lambda$). First for $\lambda
\to \infty$ we compute the $f_{n}^{max}$, the difference equations
(\ref{differ}) become, \be \phi_{n} \simeq -{1\over 2} \ln
f_{n+1}f_{n-1}, ~~n=1,\ldots,\nu-3
 \label{differ1},
\ee this system has been solved (see e.g. \cite{BT}, \cite{BR})
with the solution being  (note again that the inhomogeneity term is omitted),\be f_{n}^{max} ={\sin ^{2}({\pi \over
\nu}) \over \sin ^{2}({\pi (n+1) \over \nu})}, ~~n=1, \ldots, \nu
-3. \label{sol1} \ee Similarly, for $\lambda \to -\infty$ \be
\phi_{n} &\simeq& -{1\over 2} \ln f_{n+1}f_{n-1}, ~~n=1, \ldots,
\nu -3, ~~n\neq q_{1}, q_{2}
 \non\\ \phi_{q_{1}} & \to & -\infty ,\phi_{q_{2}}  \to  -\infty  \label{differ2},
\ee the solution of the later system has the following form \be f_{n}^{min} &=& {\sin
^{2}({\pi \over q_{2} +2}) \over \sin ^{2}({\pi (n+1) \over
q_{2} +2})}, ~~n=1, \ldots, q_{2}-1, ~~f_{q_{2}}^{min} =1 \non\\
f_{n}^{min} &=& {\sin ^{2}({\pi \over q_{1} - q_{2} + 2}) \over
\sin ^{2}({\pi (n-q_{2}+1) \over q_{1}-q_{2} +2})},
~~n=q_{2}+1, \ldots, q_{1}-1, ~~f_{q_{1}}^{min} =1 \non\\
f_{n}^{min} &=& {\sin ^{2}({\pi \over \nu-q_{1}}) \over \sin
^{2}({\pi (n - q_{1} +1) \over \nu -q_{1}})}, ~~n=q_{1}+1, \ldots,
\nu-3. \label{sol2} \ee Notice that the main difference with the
corresponding solution in \cite{BR} is the appearance of the
middle term in (\ref{sol2}) (for $n=q_{2}+1,\ldots ,q_{1}-1$), in
\cite{BR} there is no such term in the solution since
$q_{1}=q_{2}=p$. According to equation (\ref{entropy1}) and the
above solutions, the entropy can be written as \be s &=& -{4T
\over \pi}  \sum_{n=2}^{\nu- 2}\Big \{ L({\sin ^{2}({\pi \over
\nu}) \over \sin ^{2}({\pi n \over \nu})})
-\sum_{n=2}^{q_{2}}L({\sin ^{2}({\pi \over q_{2} +2}) \over \sin
^{2}({\pi
n \over q_{2} +2})}) -2L(1) \non\\
&-&\sum_{n=2}^{q_{1}-q_{2}}L({\sin ^{2}({\pi \over q_{1}-q_{2}
+2}) \over \sin ^{2}({\pi n \over q_{1}-q_{2} +2})})-\sum_{n=2}^{
\nu-q_{1}-2}L({\sin ^{2}({\pi \over \nu-q_{1} }) \over \sin
^{2}({\pi n \over \nu-q_{1}})})\Big \}.\ee  Moreover,  \be
\sum_{n=2}^{q-2}L({\sin ^{2}({\pi \over q}) \over \sin ^{2}({\pi n
\over q})}) = {2(q-3) \over q}L(1), ~~q>3 \label{i2} \ee and $L(1)
={\pi ^2 \over 6}$ (see e.g. \cite{BR}), then \be s = { 2 \pi T
\over 3} \Big ( {3q_{2} \over q_{2} +2} +{3 \delta q \over \delta
q +2}-{6q_{1} \over\nu(\nu-q_{1})}\Big ). \label{s1} \ee The
knowledge of the entropy allows the calculation of the heat
capacity, in particular \be C_{u} = T{\partial s(T) \over \partial
T} = -T {\partial^{2} f(T) \over
\partial^{2} T}, \label{c} \ee
also, at low temperature it has been shown that \cite{bc},
\cite{af}, \be C_{u} = {\pi c \over 3u }T +... \label{cap} \ee
where $c$ is the central charge of the effective conformal field
theory, and $u$ is the speed of sound (Fermi velocity).
 By means of  (\ref{s1}), (\ref{c}) and (\ref{cap})
($u={1 \over 2}$ in our notation, see e.g. \cite{FT2}) we can
readily
deduce the central charge \be c={3q_{2} \over q_{2}+2} +{3 \delta
q \over \delta q +2}-{6q_{1} \over\nu(\nu-q_{1})}. \label{central}
\ee Recall the LR symmetry condition $q_{1} =\nu -2 -q_{2}$, then
the conformal anomaly can be expressed in terms of $q_{2}$ and
$\nu$ as \be c={3q_{2} \over q_{2}+2} -{6q_{2}
\over\nu(\nu-q_{2})}+{3q_{2} \over q_{2}+2} -{6q_{2} \over \tilde
\nu(\tilde \nu-q_{2})}, \label{central1} \ee where $\tilde \nu =
\nu -q_{2}$. Note that the later expression is written in terms of
the central charges of two copies of the generalized $SU(2)$
diagonal coset theory. More
specifically, the conformal anomaly (\ref{central1}) is identified
as the sum of the central charges of the ${\cal M}(q_{2},\nu
-q_{2} -2)$ and ${\cal M}(q_{2},\tilde \nu -q_{2} -2)\equiv{\cal
M}(q_{2},\delta q)$ coset models, therefore the effective
conformal field theory should be of the form ${\cal M}(q_{2},\nu
-q_{2} -2) \otimes {\cal M}(q_{2},\delta q)$.

Expression (\ref{central}) for $q_{1}=q_{2}$ is compatible with the
result obtained by Bazhanov and Reshetikhin
---in the ``anti--ferromagnetic'' regime\footnote {the analysis of the ``ferromagnetic'' regime
is exactly the same as in \cite{BR}, and it gives rise to the
central charge of the parafermionic ${SU(2)_{\nu -2} \over U(1)}$
theory i.e., $c= 2-{6\over \nu}$, \cite{FZ}.}--- in \cite{BR}. In
the special case where $q_{2} =1$, the central charge becomes \be
c= 2 -{12 \over \nu (\nu-2)} = 1- {6 \over \nu (\nu -1)} +1- {6
\over (\nu -1)(\nu -2)} \ee and it agrees with the $c_{IR}$
presented in \cite{rs}, given by the sum of the central charges of
two unitary minimal models. Finally, in the isotropic limit the
central charge (\ref{central}) reduces to the one of the
alternating ${q_{1}\over 2}$, ${q_{2}\over 2}$ quantum spin chain
(see e.g. \cite{AM}, \cite{bydo}).

\section{Discussion}

The thermodynamics of the critical $RSOS(q_{1}, q_{2}; q)$ model,
obtained by fusion, was studied and the high and low temperature
expansion were discussed. The main result of this work was the
derivation of the effective conformal anomaly (\ref{central}),
(\ref{central1}) of the model, the validity of which was confirmed
by various tests. More specifically, for $q_{2} =1$ expression
(\ref{central1}) coincides with the $c_{IR}$ presented in
\cite{rs}, and it is specified by the sum of the central charges
of the unitary minimal models ${\cal M}_{\nu}$, ${\cal M}_{\nu
-1}$, where \be c=1-{6 \over \nu(\nu-1)} \ee is the central charge
of the unitary minimal model ${\cal M}_{\nu}$ of conformal field
theory \cite{bpz}. Also, in the case where $q_{1}=q_{2}$ we
recover the results of \cite{BR}. Finally, in the isotropic limit
$\nu \to \infty$ our result agrees with the conjectured central
charge for the alternating spin chain \cite{AM}, expressed as the
sum of the central charges of $SU(2)_{q_{2}}$, $SU(2)_{\delta q}$,
i.e., \be c={3q_{2} \over q_{2} +2} +{3 \delta q \over \delta q
+2}. \ee An exact calculation of the effective central charge for
the alternating spin chain, by means of the finite size effects
and the thermodynamic Bethe ansatz analysis, is presented in
\cite{bydo}. In general, the central charge (\ref{central1})
obtained in the present study is identified as the sum of the
central charges of the ${\cal M}(q_{2},\nu -q_{2} -2)$ and ${\cal
M}(q_{2},\delta q)$ coset models, whereas in \cite{BR}
Bazhanov and Reshetikhin by studying the $RSOS(p,q)$ models found an effective central charge that
corresponds to the ${\cal M}(p,\nu -p -2)$ model. We conclude that
the effective conformal field theory that emanates from the study
of the $RSOS(q_{1}, q_{2};q)$ model, consists of two copies of the
generalized $SU(2)$ coset theory.

A compelling task is to extend the above calculations in the
presence of boundaries, and compute the boundary energy of the
system as well as the corresponding $g$--function (see e.g.
\cite{al}--\cite{drtw}). There exist solutions of the boundary
Yang--Baxter equation \cite{cherednik} in the $RSOS$
representation \cite{rsosb1}--\cite{rsosb3}, and moreover, in
\cite{rsosb1} the Bethe ansatz equations of the $RSOS$ model with
boundaries have been explicitly derived. Finally, a very
challenging problem is the formulation of a string hypothesis for
integrable critical models associated with non--simply laced
algebras such as the $A_{2}^{(2)}$ (Izergin--Korepin) quantum spin
chain \cite{ik}. Such a formulation is necessary for the
investigation of the thermodynamics as well as the conformal
properties of these systems.

\vspace{.25in}

 {\large \bf Acknowledgments} I am indebted  to A. Babichenko for
 useful discussions, especially on the interpretation of the structure of
 the conformal anomaly (\ref{central1}),
 and for prior collaboration. I also would like to thank  EPSRC
 for a research fellowship.

\end{document}